\documentclass{aastex63}
\usepackage[latin9]{inputenc}
\usepackage{array}
\usepackage{amsbsy}
\usepackage{amstext}
\usepackage{graphicx}

\makeatletter

\providecommand{\tabularnewline}{\\}

\usepackage{array}
\usepackage{amsbsy}
\usepackage{xcolor}

\providecommand{\tabularnewline}{\\}

\makeatother

\shorttitle{Torsional Oscillations and Helioseismic Predictions}
\shortauthors{V. Pipin and A. Kosovichev}

\begin{document}

\title{Torsional Oscillations in Dynamo Models with Fluctuations 
and Potential for Helioseismic Predictions of the Solar Cycles}

\author{Valery V. Pipin}
\affiliation{Institute of Solar-Terrestrial Physics, Russian Academy of
Sciences, Irkutsk, 664033, Russia}
\author{Alexander G. Kosovichev}
\affiliation{Center for Computational Heliophysics, New Jersey Institute of Technology, Newark, NJ 07102, USA}
\affiliation{Department of Physics, New Jersey Institute of Technology, Newark, NJ 07102, USA}
\affiliation{NASA Ames Research Center, Moffett Field, CA 94035, USA}
\begin{abstract}
Using a nonlinear mean-field solar dynamo model, we study relationships
between the amplitude of the `extended' mode of migrating zonal
flows (`torsional oscillations') and magnetic cycles, and investigate
whether properties the torsional oscillations in subsurface layers and in the deep convection zone can provide information about the future solar cycles.
We consider two types of dynamo models: models with regular variations
of the alpha-effect, and models with stochastic fluctuations, { simulating `long'- and 'short-memory' types of magnetic activity variations.
It is found that torsional oscillation parameters, such the zonal acceleration, show a considerable correlation with the magnitude of the subsequent cycles with a time lag of 11-20 yr. The sign of the correlation and the time-lag parameters can depend on the depth and latitude of the torsional oscillations as well as on the properties of long-term (`centennial') variations of the dynamo cycles. The strongest correlations are found for the zonal acceleration at high latitudes at the base of the convection zone. The model results demonstrate that helioseismic observations of the torsional oscillations can be useful for advanced prediction of the solar cycles, one-two sunspot cycles ahead.}   
\end{abstract}

\section{Introduction}

According to the current knowledge, global hydromagnetic dynamo acting
inside the Sun determines the nature of the solar magnetic activity.
Parker \citeyearpar{Parker1955} showed that the dynamo action involves
the cyclic transformation of poloidal and toroidal components of the
global magnetic field of the Sun. This scenario suggests that the
magnetic field of bipolar active regions is formed from the large-scale
toroidal magnetic field that is generated from the axisymmetric poloidal
magnetic field by differential rotation deep in the convection zone.
Parker \citeyearpar{Parker1955} and \citet{Krause1980} suggested
a mechanism of generation of the large-scale poloidal magnetic field
from the toroidal field through a turbulent electromotive force
excited by cyclonic convection. It is the so-called `alpha - effect'.
The mechanism of the generation of the large-scale poloidal magnetic
field is not yet fully established. Several alternative mechanisms
of the poloidal field generation can be found in the literature \citep{Babcock1961,Choudhuri1999,Charbonneau2011,Cameron17}.
In general, the dynamo action provides mutual cyclic amplification
of the poloidal and toroidal components of the large-scale magnetic
field (LSMF). In a stationary regime, the dynamo generation saturates
due to nonlinear effects, e.g., because of magnetic helicity conservation
(\citealp{Kleeorin1999,Kleeorin2000}), magnetic buoyancy \citep{Parker1984},
and magnetic feedback on the angular momentum and heat transport in
the solar convection zone (see reviews of \citealp{Brandenburg2005b,Brandenburg2018}).

A well-known method of solar-cycle forecast employs an empirical relationship
between the amplitude of the generated toroidal field and the strength
of the poloidal magnetic field during the preceding solar minima \citep{Schatten1978}.
This relationship makes it possible to predict the sunspot maxima
from the amplitude of the polar field observed during the previous
solar activity minima. The forecast horizon of this method is approximately 5-6 years, half the 11-year solar cycle. This relationship is employed in the
flux-transport and Babcock-Leighton types of dynamo models \citep{Choudhuri2007}.
One possibility to improve the forecast is to take into account the
nonlinear relationship between global flows and magnetic fields of
the Sun. From the analysis of helioseismic measurements of zonal flows
migrating during the solar cycle in the convection zone (so-called
`torsional oscillations') \citet{Kosovichev2019} argued that
the amplitude of the zonal acceleration in a high-latitude region
at the base of the convection zone during the solar
maxima may give information about the strength of the following sunspot maxima.
{ If confirmed,} this relationship { would} give to the solar activity forecast a full 11-year cycle ahead.

We study theoretical relationships between the variations
of the torsional oscillations and the amplitude of solar cycles. In particular, 
we consider fully dynamical mean-field dynamo models that reproduce 
the observed `extended' 22-year mode of the torsional oscillations.
The extended wave of zonal variations of rotation propagates from
high latitudes to the equator during the 22-year `extended' solar
cycle \citep{Altrock1997,Ulrich2005}. Recently, \citet{Kosovichev2019}
and \citet{Pipin2019c} presented observational and theoretical evidence in favor of the global nature of this wave in the solar dynamo process.
{ These results indicated a possibility for using characteristics of
the torsional oscillations, inferred through helioseismology analysis, for solar-cycle forecasting. The advantage of this approach is that helioseismology can provide measurements of the torsional oscillations through the whole convection zone, including the tachocline region which is a primary seat of the solar dynamo.}

In this study, we employ the non-linear dynamo model that couples the
magnetic field evolution with global dynamic and thermodynamic variations
in the convection zone and provides a realistic description of the
torsional oscillations and their extended mode.
{ To investigate correlations between the torsional oscillations and
magnetic activity, we consider variations of the magnetic activity cycles, which are modeled using fluctuations of the alpha-effect. We consider two types: a long-term evolution corresponding to `centennial' variations of the solar activity, and short-term random fluctuations.}  In Section~2, we describe the models. In Section~3,
we present results for a series of dynamo models { and investigate 
correlations between the subsurface and surface flow characteristics and
the subsurface toroidal magnetic field of the subsequent activity cycles}. 
We evaluate the forecasting potentials and compare them with the correlations based on the relationship of the polar field strength and the following cycle amplitude. 
The paper concludes with a discussion of the main results.

\section{Basic equations}

\subsection{Dynamo model}

A detailed description of the dynamo model can be found in our previous
paper \citep[ hereafter PK19]{Pipin2019c}. The model describes the
dynamo generation of large-scale magnetic fields (LSMF) in the bulk
of the solar convective zone (CZ). The model is based on the mean-field
induction equation \citep{Krause1980}:
\begin{equation}
\mathrm{\partial_{t}\overline{\mathbf{B}}=\boldsymbol{\nabla}\times\left(\boldsymbol{\mathcal{E}}+\mathbf{\overline{U}}\times\overline{\mathbf{B}}\right)},\label{eq:mfe-1}
\end{equation}
where the induction vector of the LSMF, $\overline{\mathbf{B}}$,
is represented as the sum of the toroidal and poloidal components:
\[
\overline{\mathbf{B}}=\hat{\mathbf{\boldsymbol{\phi}}}B+\nabla\times\frac{A\hat{\mathbf{\boldsymbol{\phi}}}}{r\sin\theta},
\]
where $r$ is the radial distance, $\theta$ is the polar angle, $\hat{\mathbf{\boldsymbol{\phi}}}$
is the unit vector in the azimuthal direction. 
The mean electromotive force $\boldsymbol{\mathcal{E}}$ describes
the turbulent generation effects, pumping, and diffusion: 
\begin{equation}
\mathcal{E}_{i}=\left(\alpha_{ij}+\gamma_{ij}\right)\overline{B}_{j}-\eta_{ijk}\nabla_{j}\overline{B}_{k}.\label{eq:EMF-1-1}
\end{equation}
where the symmetric tensor $\alpha_{ij}$ stands for the turbulent
generation of the LSMF by kinetic and magnetic helicities; the antisymmetric
tensor $\gamma_{ij}$ describes the turbulent pumping effect; the
anisotropic (in the general case) tensor $\eta_{ijk}$ is the eddy
diffusivity of the LSMF \citet{Pipin2018b}. The large-scale (LS)
flow field, $\mathbf{\overline{U}}=\mathbf{\overline{U}}^{m}+r\sin\theta\Omega\left(r,\theta\right)\hat{\mathbf{\boldsymbol{\phi}}}$
produces the LS toroidal magnetic field from the LS poloidal field
by means of the differential rotation, $\Omega\left(r,\theta\right)$.
The meridional circulation, $\mathbf{\overline{U}}^{m}$, advects
the LSMF in the convection zone. The angular momentum conservation
and the equation for the azimuthal component of large-scale vorticity,
$\mathrm{\overline{\omega}=\left(\boldsymbol{\nabla}\times\overline{\mathbf{U}}^{m}\right)_{\phi}}$,
determine distributions of the differential rotation and meridional
circulation: 
\begin{eqnarray}
\frac{\partial}{\partial t}\overline{\rho}r^{2}\sin^{2}\theta\Omega & = & -\boldsymbol{\nabla\cdot}\left(r\sin\theta\overline{\rho}\left(\hat{\mathbf{T}}_{\phi}+r\sin\theta\Omega\mathbf{\overline{U}^{m}}\right)\right)\label{eq:angm}\\
 & + & \boldsymbol{\nabla\cdot}\left(r\sin\theta\frac{\overline{\mathbf{B}}\overline{B}_{\phi}}{4\pi}\right),\nonumber 
\end{eqnarray}

\begin{eqnarray}
\mathrm{\frac{\partial\omega}{\partial t}\!\!\!} & \mathrm{\!\!=\!\!\!\!} & \mathrm{r\sin\theta\boldsymbol{\nabla}\cdot\left(\frac{\hat{\boldsymbol{\phi}}\times\boldsymbol{\nabla\cdot}\overline{\rho}\hat{\mathbf{T}}}{r\overline{\rho}\sin\theta}-\frac{\mathbf{\overline{U}}^{m}\overline{\omega}}{r\sin\theta}\right)}\label{eq:vort}\\
 & + & \mathrm{r}\sin\theta\frac{\partial\Omega^{2}}{\partial z}-\mathrm{\frac{g}{c_{p}r}\frac{\partial\overline{s}}{\partial\theta}}\nonumber \\
 & + & \frac{1}{4\pi\overline{\rho}}\left(\overline{\mathbf{B}}\boldsymbol{\cdot\nabla}\right)\left(\boldsymbol{\nabla}\times\overline{\mathbf{B}}\right)_{\phi}-\frac{1}{4\pi\overline{\rho}}\left(\left(\boldsymbol{\nabla}\times\overline{\mathbf{B}}\right)\boldsymbol{\cdot\nabla}\right)\overline{\mathbf{B}}{}_{\phi},\nonumber 
\end{eqnarray}
where $\hat{\mathbf{T}}$ is the turbulent stress tensor: 
\begin{equation}
\hat{T}_{ij}=\left(\left\langle u_{i}u_{j}\right\rangle -\frac{1}{4\pi\overline{\rho}}\left(\left\langle b_{i}b_{j}\right\rangle -\frac{1}{2}\delta_{ij}\left\langle \mathbf{b}^{2}\right\rangle \right)\right),\label{eq:rei}
\end{equation}
(see detailed description in PK19). Also, $\overline{\rho}$ is the
mean density, $\mathrm{\overline{s}}$ is the mean entropy; $\mathrm{\partial/\partial z=\cos\theta\partial/\partial r-\sin\theta/r\cdot\partial/\partial\theta}$
is the gradient along the axis of rotation. The mean heat transport
equation determines the mean entropy variations from the reference
state due to the generation and dissipation of LSMF and large-scale flows
\citep{Pipin2000}: 
\begin{equation}
\overline{\rho}\overline{T}\left(\frac{\partial\overline{\mathrm{s}}}{\partial t}+\left(\overline{\mathbf{U}}\cdot\boldsymbol{\nabla}\right)\overline{\mathrm{s}}\right)=-\boldsymbol{\nabla}\cdot\left(\mathbf{F}^{c}+\mathbf{F}^{r}\right)-\hat{T}_{ij}\frac{\partial\overline{U}_{i}}{\partial r_{j}}-\boldsymbol{\boldsymbol{\mathcal{E}}}\cdot\left(\nabla\times\overline{\boldsymbol{B}}\right),\label{eq:heat}
\end{equation}
where $\overline{T}$ is the mean temperature, $\mathbf{F}^{r}$ is
the radiative heat flux, $\mathbf{F}^{c}$ is the anisotropic convective
flux. An analytical mean-field expression for $\mathbf{F}^{c}$ takes
into account the effect of the Coriolis force, and the influence of
the LSMF on the turbulent convection (see, PK19). The last two terms
in Eq (\ref{eq:heat}) take into account the convective energy gain
and sink caused by the generation and dissipation of LSMF and large-scale
flows. The reference profiles of mean thermodynamic parameters, such
as entropy, density, and temperature are determined from the stellar
interior model MESA \citep{Paxton2011,Paxton2013}. The radial profile
of the typical convective turnover time, $\tau_{c}$, is determined
from the MESA code, as well. We assume that $\tau_{c}$ does not depend
on the magnetic field and global flows. The convective RMS velocity
is determined from the mixing-length approximation, 
\begin{equation}
\mathrm{u_{c}=\frac{\ell_{c}}{2}\sqrt{-\frac{g}{2c_{p}}\frac{\partial\overline{s}}{\partial r}},}\label{eq:uc}
\end{equation}
where $\ell_{c}=\alpha_{MLT}H_{p}$ is the mixing length, $\alpha_{MLT}=1.9$
is the mixing length parameter, and $H_{p}$ is the pressure height
scale. Eq.~(\ref{eq:uc}) determines the reference profiles for the
eddy heat conductivity, $\chi_{T}$, eddy viscosity, $\nu_{T}$, and
eddy diffusivity, $\eta_{T}$, as follows, 
\begin{eqnarray}
\chi_{T} & = & \frac{\ell^{2}}{6}\sqrt{-\frac{g}{2c_{p}}\frac{\partial\overline{s}}{\partial r}},\label{eq:ch}\\
\nu_{T} & = & \mathrm{Pr}_{T}\chi_{T},\label{eq:nu}\\
\eta_{T} & = & \mathrm{Pm_{T}\nu_{T}}.\label{eq:et}
\end{eqnarray}
The model gives the best agreement of the angular velocity profile
with helioseismology results for $\mathrm{Pr}_{T}=3/4$ (PK19). Also,
the dynamo model reproduces the solar magnetic cycle period, $\sim22$ years,
if $\mathrm{Pm}_{T}=10$.

\begin{figure}\centering
\includegraphics[width=0.95\textwidth]{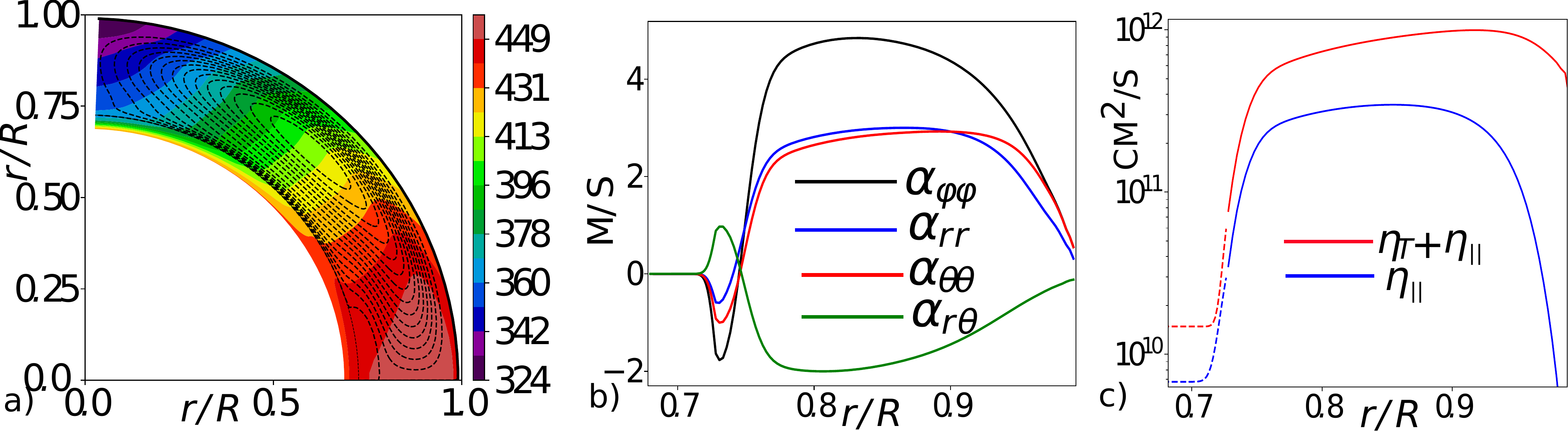} \caption{\label{fig1}
a) The basic angular velocity profile and the streamlines
of the meridional circulation; b) the radial profiles
of the $\alpha$-effect tensor at latitude 45$^{\circ}$; c) radial
profiles of the total, $\eta_{T}+\eta_{||}$, and the rotationally
induced part, $\eta_{||}$, of the eddy magnetic diffusivity. }
\end{figure}

Figure \ref{fig1} shows the angular velocity profile, streamlines
of the meridional circulation, the radial profiles of the $\alpha$
- effect, and the eddy diffusivity in the model. The magnitude of
the meridional flow on the surface is about $14$~m/s. The angular
velocity profile agrees well with helioseismology data. A detailed
theoretical discussion of the mechanisms generating the differential
rotation and meridional circulation in our model can be found in \citet{Pipin2018c}
(also see, \citealp{Kitchatinov1999a,Kitchatinov2005}). The model
includes a phenomenological description of the tachocline where the
differential rotation is transformed into a solid body rotation. We
assume that the intensity of turbulent mixing in the tachocline drops
exponentially with distance from the bottom of the convection zone
and that the $\alpha$ - effect vanishes in the tachocline. 

 { In the model, the magnetic field strength at the bottom of the 
convection zone may reach several kilogauss. The magnetic field may cause  relative variations of the convective heat flux
inside the convection zone, which can reach 2\%  \citep{Pipin2018b}.
However, these variations deep inside the convection zone do not 
cause significant changes of the heat flux on the surface, because
their effect is screened due to the huge heat capacity of 
the turbulent plasma in the solar interior \citep[e.g.][]{Stix1981,2000SSRv...94..113S}.
In the models presented in this paper, the maximum surface variations of the entropy do not exceed $5\times 10^{-4}$. Such variations are in line with the observed changes of the total solar irradiance.}

\subsection{Magnetic helicity and $\alpha$-effect fluctuations}

Similar to our previous papers, we employ the $\alpha$-effect tensor,
which represents a combination of kinetic and magnetic helicities
in the following form: 
\begin{eqnarray}
\alpha_{ij} & = & C_{\alpha}\left(1+\xi^{\left(\alpha\right)}\left(t\right)\right)\psi_{\alpha}(\beta)\alpha_{ij}^{(H)}+\alpha_{ij}^{(M)}\psi_{\alpha}(\beta)\frac{\overline{\chi}\tau_{c}}{4\pi\overline{\rho}\ell^{2}},\label{alp2d}
\end{eqnarray}
where $\xi^{\left(\alpha\right)}\left(t\right)$ is the fluctuating
part of kinetic helicity tensor, $\alpha_{ij}^{(H)}$ (Fig.~\ref{fig1}b
shows the radial profile of the tensor components); $\overline{\chi}=\overline{\mathbf{a\cdot b}}$
is the magnetic helicity density ($\mathbf{a}$ and $\mathbf{b}$
are the turbulent parts of the magnetic vector potential and magnetic
field vector), and tensor $\alpha_{ij}^{(M)}$ takes into account
the effect of the Coriolis force. Function $\psi_{\alpha}(\beta)$
stands for the `algebraic' saturation of the $\alpha$- effect
caused by the small-scale Lorentz force, which opposes convective motions
across the field lines of the LSMF, where, $\mathrm{\beta=\left|\overline{\mathbf{B}}\right|/\sqrt{4\pi\overline{\rho}u_{c}^{2}}}$.
For strong LSMF, when $\beta\gg1$, $\psi_{\alpha}(\beta)\sim\beta^{-3}$.
A detailed description of $\alpha_{ij}^{(H)}$, $\alpha_{ij}^{(M)}$
and $\psi_{\alpha}(\beta)$ is given by \citet{Pipin2018b}. The magnetic
helicity evolution follows the conservation law: 
\begin{equation}
\frac{\partial\overline{\chi}^{(tot)}}{\partial t}=-\frac{\overline{\chi}}{R_{m}\tau_{c}}-2\eta\overline{\mathbf{B}}\cdot\mathbf{\overline{J}}-\boldsymbol{\nabla\cdot}\boldsymbol{\boldsymbol{\text{F}}}-\mathbf{\left(\overline{U}\cdot\boldsymbol{\nabla}\right)}\overline{\chi}^{(tot)}\label{eq:helcon}
\end{equation}
where 
\begin{equation}
\overline{\chi}^{(tot)}=\overline{\mathbf{A}\cdot\mathbf{B}}=\overline{\mathbf{A}}\cdot\overline{\mathbf{B}}+\overline{\mathbf{a\cdot b}}\text{,}\label{chitot}
\end{equation}
$\mathbf{B}=\nabla\times\mathbf{A}$, $\overline{\mathbf{A}}$ is
the LSMF vector potential; $R_{m}$ is the magnetic Reynolds number,
(we put $R_{m}=10^{6}$). We assume that the eddy diffusivity of the magnetic
helicity is isotropic and that the diffusive helicity flux $\boldsymbol{\boldsymbol{F}}=-\eta_{\chi}\boldsymbol{\nabla}\overline{\chi}$,
where $\eta_{\chi}=0.1\eta_{T}$ \citep{Mitra2010}.

In this study, we perform several runs of the dynamo model different
parameters of the kinetic helicity. In the first three runs, models
C1, C2, and C3, we vary the dimensionless $\alpha$-effect parameter,
$C_{\alpha}=\left\{ 0.04,0.05,0.06\right\} $. The value $C_{\alpha}=0.04$
is slightly above the dynamo instability threshold. 

\begin{figure}	\centering
\includegraphics[width=0.5\columnwidth]{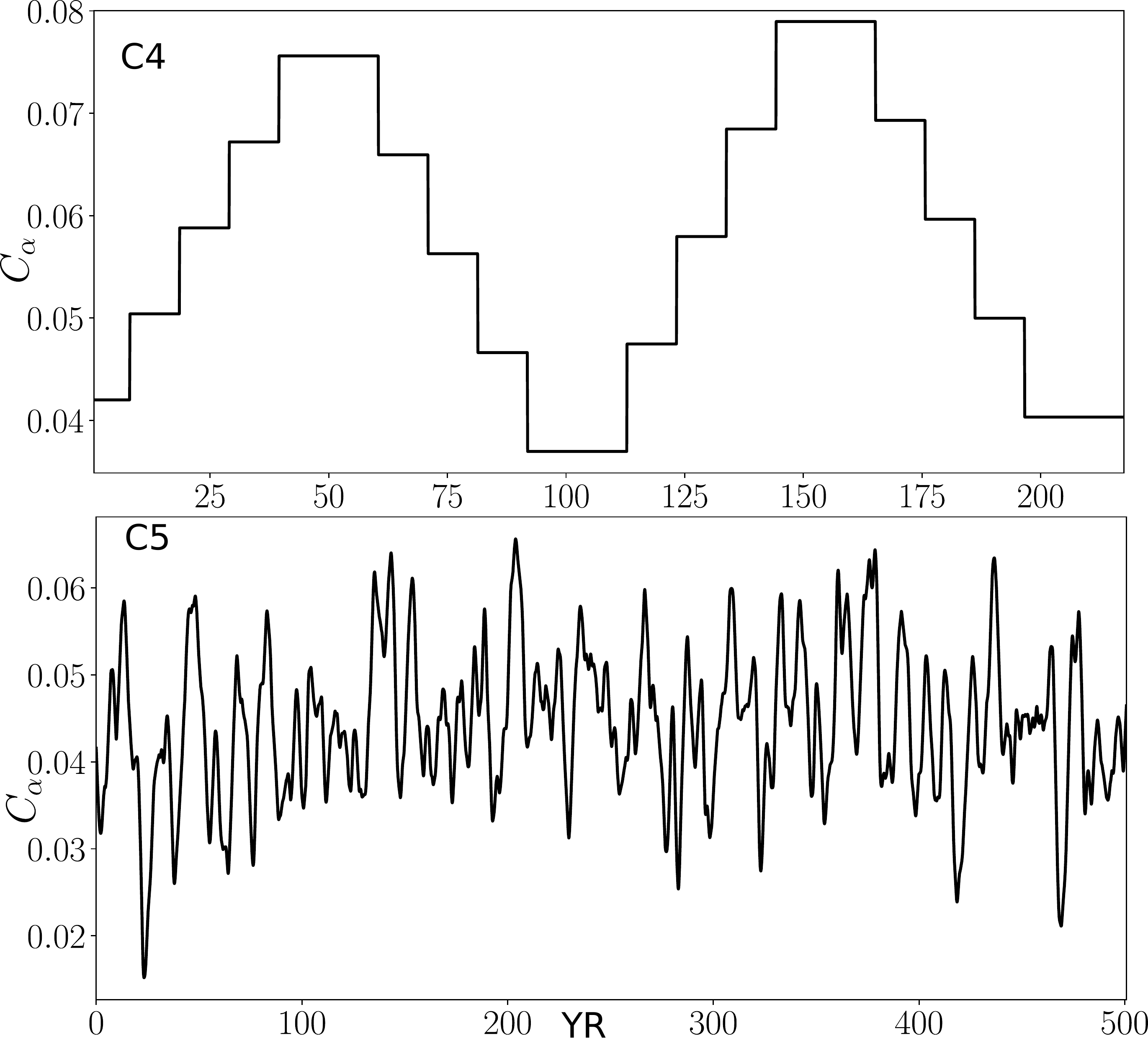} \caption{\label{fig:alf}Variations of the $\alpha$-effect parameter, $C_{\alpha}$
		in Models C4 and C5}
\end{figure}

In model C4, we simulate long-term magnetic activity variations 
by increasing $C_{\alpha}$
from 0.04 to 0.08 after each cycle and then decreasing it back in
the same way. {In model C4, the $\alpha$ - effect coefficient,
$C_{\alpha}$, increases by a constant value after each half-cycle
and then drops below the critical threshold (Fig.~\ref{fig:alf}a).
It was carried out as follows. We start the run with the magnetic
field distribution taken around the magnetic cycle minimum and
with the slightly super-critical $C_{\alpha}=0.04$. Then, 
every ten years, we increase  $C_{\alpha}$ by $0.01$ until it reaches
$0.08$. After this moment, we decrease the parameter every ten years by
the same amount $-0.01$ until the value $C_{\alpha}=0.02$. Then, the
procedure is repeated. } 

In model C5, we consider random fluctuations of the $\alpha$-effect in time (Fig.~\ref{fig:alf}b), which simulate random variations of the cycle amplitude.
Similarly to \citet{Rempel2005c} and \citet{Mordvinov2018}, we model
the random parameter $\xi^{\left(\alpha\right)}$ via the Ornstein--Uhlenbeck
process, i.e., the evolution of the $\xi^{\left(\alpha\right)}$ is
governed by the systems of the stochastic differential  equations,
\begin{eqnarray*}
\dot{\xi}^{\left(\alpha\right)} & = & -\frac{2}{\tau_{\xi}}\left(\xi^{\left(\alpha\right)}-\xi_{1}\right),\\
\dot{\xi}_{1} & = & -\frac{2}{\tau_{\xi}}\left(\xi_{1}-\xi_{2}\right),\\
\dot{\xi}_{2} & = & -\frac{2}{\tau_{\xi}}\left(\xi_{2}-g\sqrt{\frac{2\tau_{\xi}}{\tau_{h}}}\Theta\right),
\end{eqnarray*}
where $g$ is a Gaussian random number which is renewed at every time
step, $\tau_{h}$ is the time step of the numerical simulations, $\tau_{\xi}$
is the relaxation time of $\xi^{\left(\alpha\right)}$ and $\xi_{1,2,3}$
are auxiliary parameters that are introduced to smooth variations
of $\xi^{\left(\alpha\right)}$ with its first- and second-order derivatives.
To generate the $\alpha$-effect randomness in colatitude we introduce
the random function $\Theta$, renewed at each time step as well.
It is defined as follows. We generate spatially random Gaussian sequences,
$\Theta\left(\theta_{j}\right)$, where $\theta_{j}$ are the collocation
points of the Legendre polynomials, and $\langle\Theta\left(\theta_{j}\right)\rangle=0$,
$\sigma\left(\Theta\right)=1$. Then, the sequence $\Theta\left(\theta_{j}\right)$
is decomposed into the Legendre polynomials. Finally, we filter out
all the Legendre harmonics higher than $\ell=5$, and normalize $\Theta\left(\theta\right)$
to unity. The resulted latitudinal fluctuations of the $\alpha$-effect
are described by the smooth functions. We use a small level of 
long-term fluctuations with $\sigma\left(g\right)=0.2$ and $\tau_{\xi}=2$ yr. 

{ Models C4 and C5 can be considered as `long-' and `short-memory' type models of the solar cycle. While the exact cause of the observed solar-cycle variations is not known 
these models are suitable to study statistical relations between the torsional oscillations and the magnetic cycle parameters. Parameters of the models are given in Table \ref{tab:}.}

\begin{table}
\begin{tabular}{l>{\raggedright}p{2.6cm}>{\raggedright}p{1.5cm}p{1.2cm}>{\raggedright}p{1.8cm}>{\raggedright}p{1.9cm}}
\hline 
Model  & \noindent $C_{\alpha}$  & $P${[}YR{]}  & D{[}G{]}  & $B_{\ensuremath{\phi}}${[}G{]}  & $\partial_{t}\overline{U}_{\phi}${[}$10\ensuremath{^{-8}}$m$\,$s$^{-2}${]}\tabularnewline
\hline 
C1  & \noindent 0.04  & 12  & 3.2  & 800/2700  & 2.4\tabularnewline
C2  & \noindent 0.05  & 10.2  & 4.3  & 990/3100  & 3.4\tabularnewline
C3  & \noindent 0.065  & 8.3  & 5.0  & 1350/3500  & 5.8\tabularnewline
C4  & 0.03 - 0.08  & 7.5/14  & 3 /5.1  & 1500/3600  & 7.0\tabularnewline
C5  & \noindent 0.05$\times${\small{}{}{}$\left(1+\xi^{\left(\alpha\right)}\right)$}, $\sigma\left(\xi\right)=0.3$,
$\tau_{\xi}=2$yr  & 7.5/13  & 3.2/4.9  & 1500/3600  & 10.0\tabularnewline
\hline 
\end{tabular}\caption{\label{tab:}Model parameters: $C_{\alpha}$ is the dimensionless
parameter of the $\alpha$-effect; $P$ is half of the dynamo period; $D$ is
the maximum strength of the dipole component of the LSMF; $B_{\ensuremath{\phi}}$
is the maximum strength of the toroidal LSMF in the subsurface layer
$r=0.9\,R$, and near the bottom of the convection zone ($r=0.74\,R$);
$\partial_{t}\overline{U}_{\phi}$ is the amplitude of solar-cycle
variations of the zonal acceleration.}
\end{table}

\begin{figure}\centering
\includegraphics[width=1\textwidth]{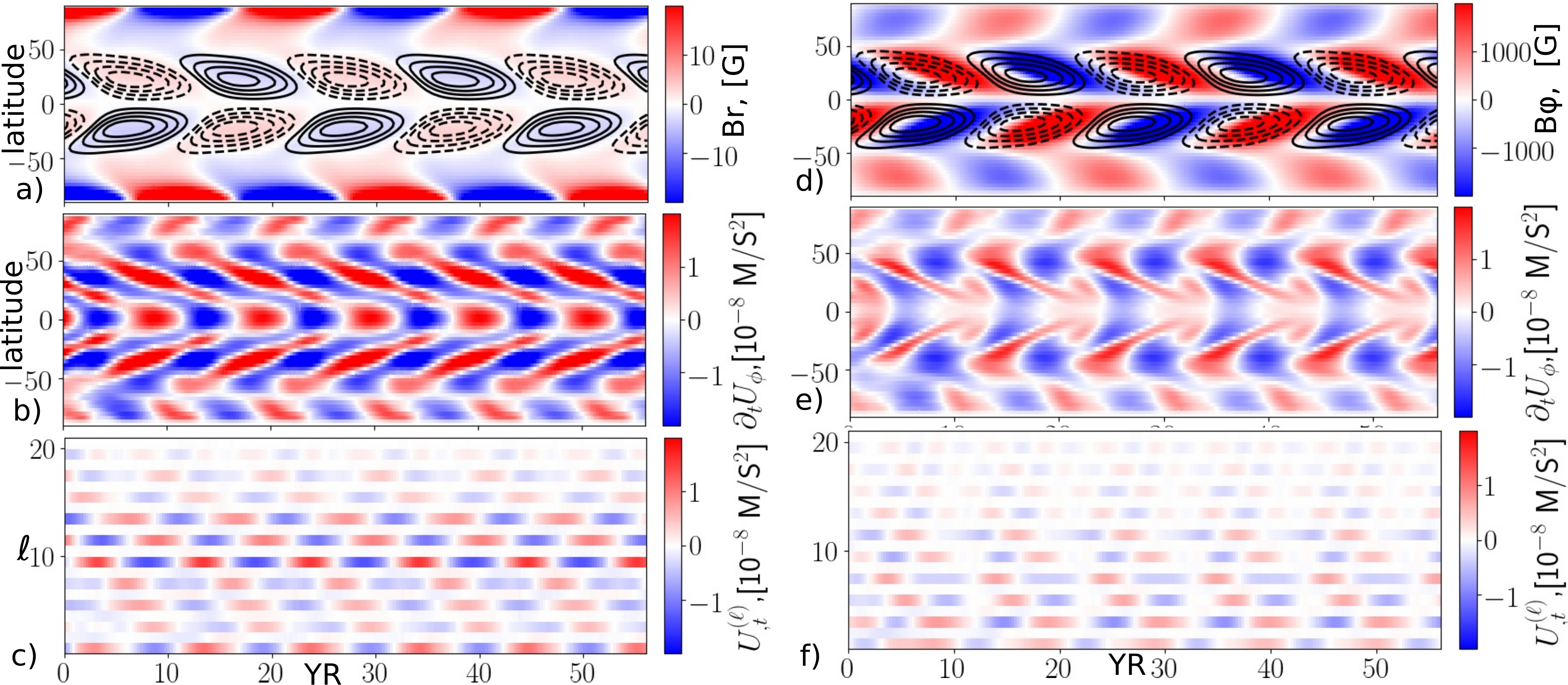}
\caption{\label{figm2} Model C2, a) variations of the radial magnetic field
at the surface (color image) and the strength of the toroidal magnetic
field at $r=0.95\,R_\odot$ (contour lines cover the interval $\pm$1 kG);
b) the zonal acceleration at the surface; c) variations of low order
harmonics $U_{,t}^{(\ell)}$ ($\ell=1-21$) of the zonal acceleration
at the surface; d) contours show the toroidal LSMF at $r=0.95R$,
and the background color image shows the toroidal LSMF evolution near
the bottom of the convection zone, $r=0.73\,R_\odot$; e) and f) the same
as in panels b) and c) for $r=0.73\,R$.}
\end{figure}

\section{Results}
\subsection{Model characteristics}

Figure \ref{figm2} shows the time-latitude diagrams evolution of
the LSMF and the zonal acceleration at the surface and the bottom
of the convection zone for model C2. The time-latitude diagrams are
similar to those published in PK19. The wave-like migration of the
toroidal LSMF has two branches: polar and equatorial. The equatorial
branch near the bottom of the convective zone goes ahead of the wave
near the surface by about half the full magnetic cycle ($\simeq11$
yr). At the surface, the extended wave of torsional oscillations starts
propagating from high latitudes to the equator. At the same time,
a new wave of the toroidal LSMF starts near the bottom of the convective
zone.

{ To characterize} the magnetic cycle { strength}, we introduce the
total unsigned magnetic flux $F_{T}$ of the toroidal LSMF in the
subsurface layer, $0,89-0.99R_\odot$: 
\begin{equation}
F_{T}=\int_{-1}^{1}\int_{0.89R_\odot}^{0.99R_\odot}\left|\overline{B}_{\ensuremath{\phi}}\right|r\mathrm{drd\mu,}\label{eq:ft}
\end{equation}
where $\mu=\cos\theta$. Similarly, we define the toroidal magnetic
fluxes for each hemisphere, $F_{T}^{N}$ and $F_{T}^{S}$. {Note,
	that $F_{T}=F_{T}^{N}+F_{T}^{S}$.  Also, we use a decomposition of the surface
radial magnetic field into a set of normalized Legendre polynomials
$P_{\ell}$ : 
\begin{equation}
\overline{B}_{\ensuremath{r}}=\sum_{\ell=1,N}B_{r}^{(\ell)}\left(t\right)P_{\ell}\left(\cos\theta\right),\label{eq:br}
\end{equation}
and define the strength of the dipole component of the radial magnetic
field, $D=B_{r}^{(1)}$.}
For characterization of the spatial structure of the torsional 
oscillations, we consider parameters of the spectrum of zonal acceleration
in the form: 
\begin{equation}
\partial_{t}\overline{U}_{\phi}=\sum U_{,t}^{(\ell)}\left(t\right)P_{\ell}^{1}\left(\cos\theta\right),\label{eq:ul}
\end{equation}
where where $P_{\ell}^{1}$ is a set of normalized associated Legendre
polynomials. 

Figures \ref{figm2}c and \ref{figm2}f show variations
of the $U_{,t}^{(\ell)}$ for the top and the bottom of the dynamo
domain. In both cases, the spectral harmonics vary with a period equal
to half the period of the dynamo cycle ($\simeq11$ years). Moreover,
at the top of the convection zone,
the harmonic $\ell=9$ shows the most significant variations among the others.
We see that the development of the cycle is accompanied by a phase
shift progressing from high- to low-order harmonics. The maximum of
the $\ell=9$ harmonic corresponds to the initiation of the extended
mode of the solar torsional oscillations at high latitudes. Near the
bottom of the convection zone, the $\ell=3$ harmonic  shows the
largest variations. At the bottom of the convection zone, the torsional
oscillation propagates to the equator for about 12 years (Fig.~\ref{figm2}e).

\begin{figure}\centering
\includegraphics[width=0.95\textwidth]{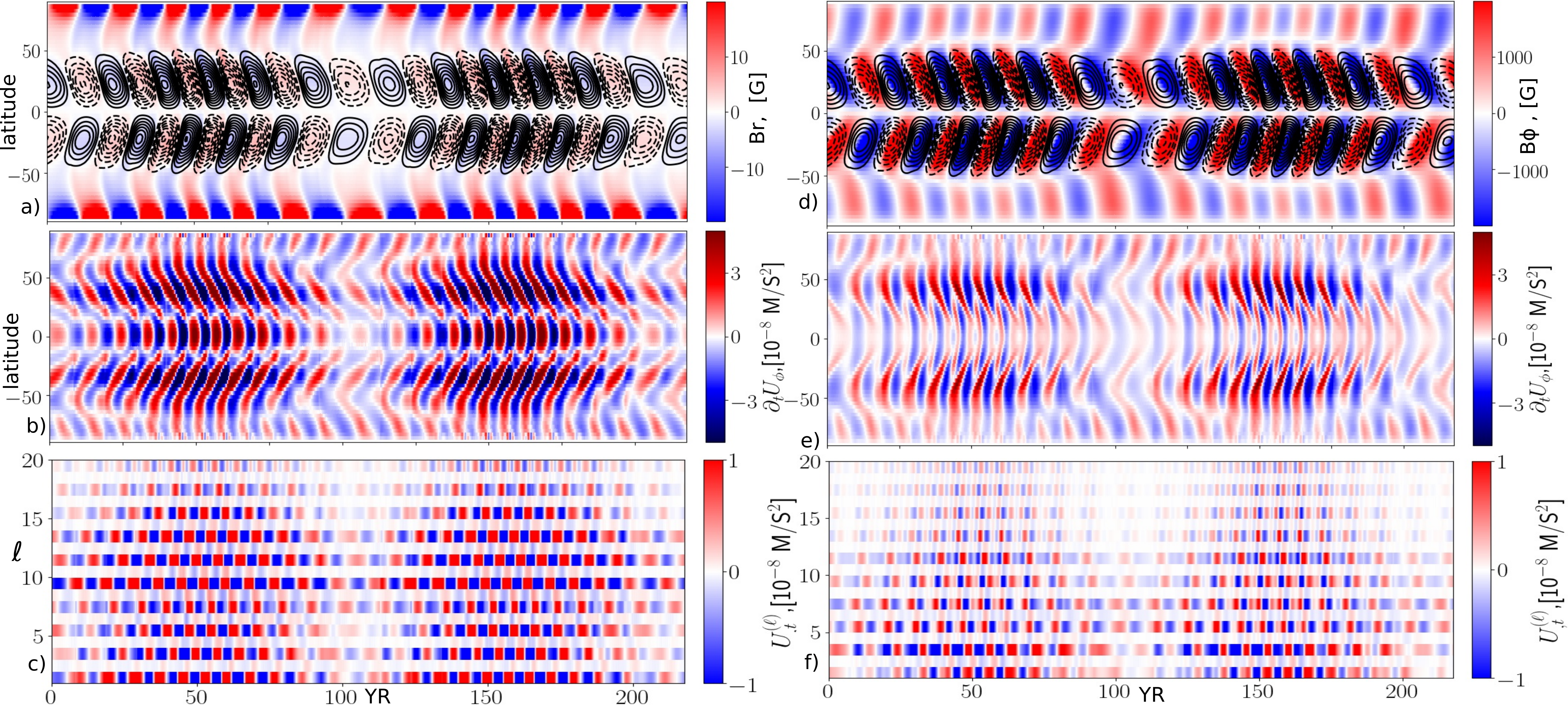} \caption{\label{fig:m5}The same as in Fig.~\ref{figm2} for Model C4.}
\end{figure}

{Next, we consider the long-term variations of the torsional
oscillations in models C4 and C5. Figure~\ref{fig:alf} shows variations
of the $\alpha$-effect parameter in these models, formulated in Sec.~2.2.}
{In Figure \ref{fig:m5}, we show the time-latitude diagrams and
the long-term evolution of the zonal acceleration modes for 
model C4.} The increase of the LSMF strength increases the magnitude
of the torsional oscillations. In the strong cycles, the polar branch
of the torsional oscillations disappears. 

\begin{figure}\centering
\includegraphics[width=0.4\textwidth]{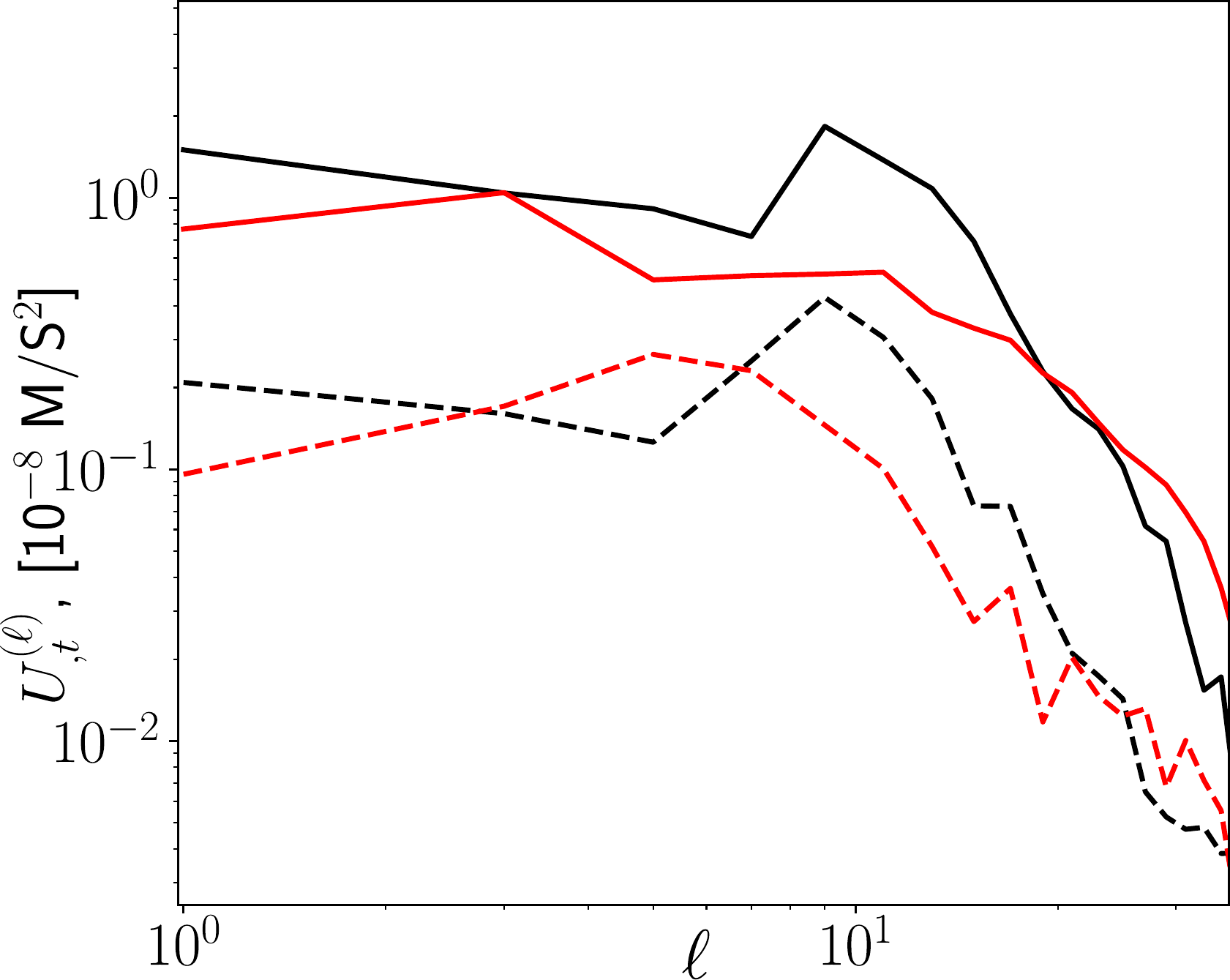}
\caption{\label{sp}Model C4, mean spectra of the torsional oscillation harmonics
of the odd order at the surface (black) and the bottom of the convection
zone (red), for the strong cycle (solid lines; the corresponding time
interval around $t=50$ yr, see Fig\ref{fig:m5}) and weak cycles
(dashed lines; the corresponding time interval around $t=100$ yr).}
\end{figure}

Figure \ref{sp} shows that, in the subsurface layer, 
the $\ell=9$ harmonic remains dominant during the periods of high
and low activity. At the bottom of the convection zone, the $\ell=3$
harmonic dominates during the long-term maxima, and the $\ell=5$
harmonic becomes stronger than the $\ell=3$ harmonic during the activity
minima. 

\begin{figure}\centering
	\includegraphics[width=1\textwidth]{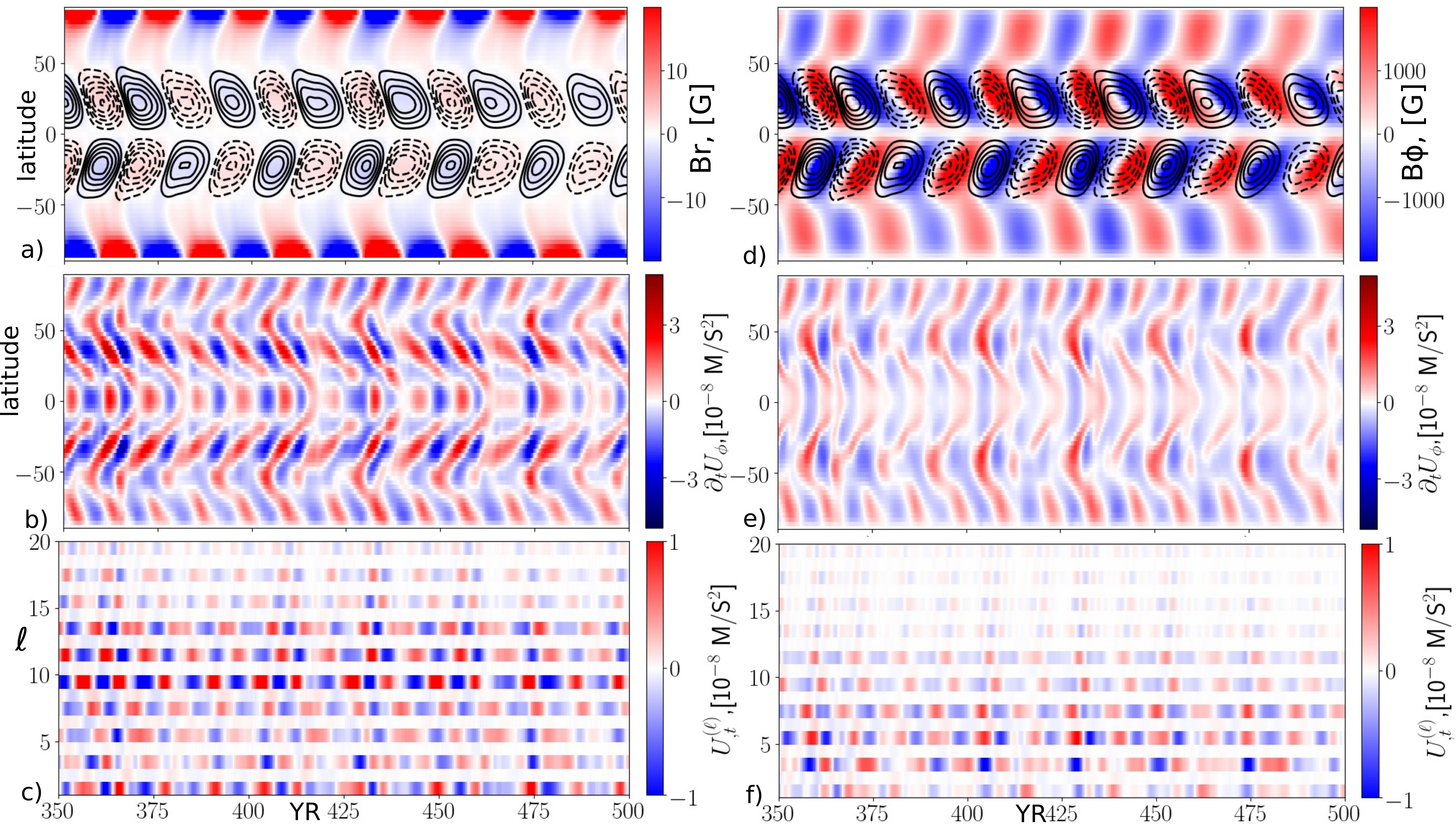}
	\caption{\label{fig:M6}The same as in Fig.~\ref{figm2} for Model C5.}
\end{figure}

{Figure \ref{fig:M6} shows the time-latitude diagrams for
	the toroidal and radial magnetic field, torsional acceleration $\partial_{t}\overline{U}_{\phi}$,
	spectral coefficients $U_{,t}^{(\ell)}$ for model C5.}
The selected time
interval includes about seven full dynamo cycles and covers about 150
years. In the subsurface layers, the strength of the toroidal LSMF  changes
from about 0.5kG during the centennial minima to 1.5 kG during the
maxima. Simultaneously, the magnitude of the total flux of the toroidal
field in the upper part of the convection zone, $F_{T}$ changes in
the range of $\left(0.4-1.2\right)\times10^{24}$ Mx. The strength
of the radial dipole LSMF is in the range of 2-5 G. The cycle duration
varies from about 8-9 years for the high amplitude cycles to 12-13
years for the weak cycles. The weakest cycle has a period of 16
years. The hemispheric asymmetry of magnetic activity in the model
is not strong. The model keeps the antisymmetric large-scale magnetic
field structure relative to the equator, which results in the dominance of the odd harmonics in
the $U_{,t}^{(\ell)}$ spectrum. Yet, we see sporadic excitation of
weak even harmonics, e.g., the harmonic of order $\ell=8$ at around
$t=360$~years in Fig.~\ref{fig:M6}c. This is caused by the deviation
of magnetic parity from the pure antisymmetric relative to the equator, 
which is quite small in the presented model.
The spectrum of the torsional oscillations at the bottom of the convection
zone in this model is qualitatively similar to model C4.

\begin{figure}\centering
	\includegraphics[width=0.8\columnwidth]{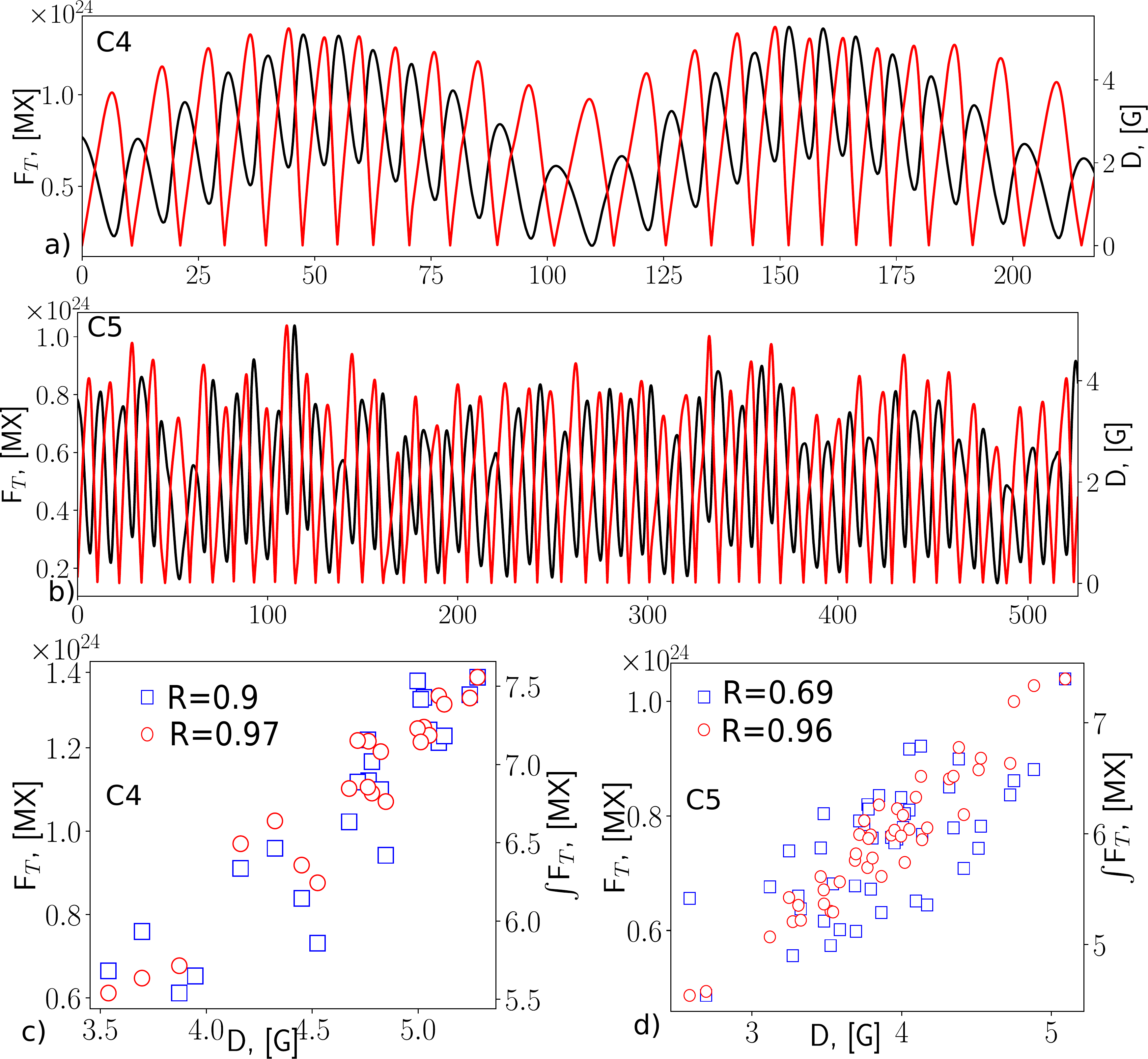}
	\caption{\label{figc5}a) Parameters $F_{T}$ (black line, left scale) and
		$D$ (red line, right scale) for model C5, b) correlation of the dipole
		magnitude in the cycle minimum with $F_{T}$ in the subsequent maximum  (blue
		squares) and the integral over the cycle $F_{T}$ (red circles) in model C5; c) the same
		as b) for model C4}
\end{figure}

\subsection{Correlations between the dipole moment and toroidal fluxes}

{ Figure \ref{figc5}
shows parameters $F_{T}$ and $D$ and their cross-correlation for different
phases of the magnetic cycles for models C4 and C5.
Figures \ref{figc5}a-b show the activity cycle of the toroidal
magnetic flux, $F_{T}$, and variations of the magnitude } of the dipole
component of the radial magnetic field, $D$, for models C4 and C5. These properties reflect two basic dynamo processes:
generation of the large-scale toroidal magnetic field by the differential
rotation and generation of the poloidal magnetic field by the helical turbulence.  Phases of the $F_{T}$ and $D$ cycles are shifted
by $\pi/2$. This means that the dipole magnitude, $D$, reaches its maximum
at the minimum of the toroidal flux, $F_{T}$, as this is observed in the solar cycles. 

Figures \ref{figc5}c-d show the correlations between the $D$ magnitude in the cycle minima with the magnitude of $F_{T}$ 
during the maxima of the subsequent cycle, for models
C4 and C5. This correlation corresponds to the well-known empirical relation between the polar magnetic field strength and the magnitude of the following sunspot cycle \citep{Schatten1978}. The quasi-regular model C4 has a higher correlation coefficient (R$\approx 0.9$)  than model C5 produced with random fluctuations of the $\alpha$-effect (R$\approx 0.69$).
Also, both models C4 and C5  have a strong correlation of $D$ in the magnetic
cycle minimum with the toroidal flux $F_{T}$ integrated over the subsequent cycle (R$\approx 0.97$ and 0.95, respectively). 

\subsection{Correlations between the toroidal oscillations and the magnetic cycle magnitude}
 
\begin{figure}\centering
\includegraphics[width=0.7\textwidth]{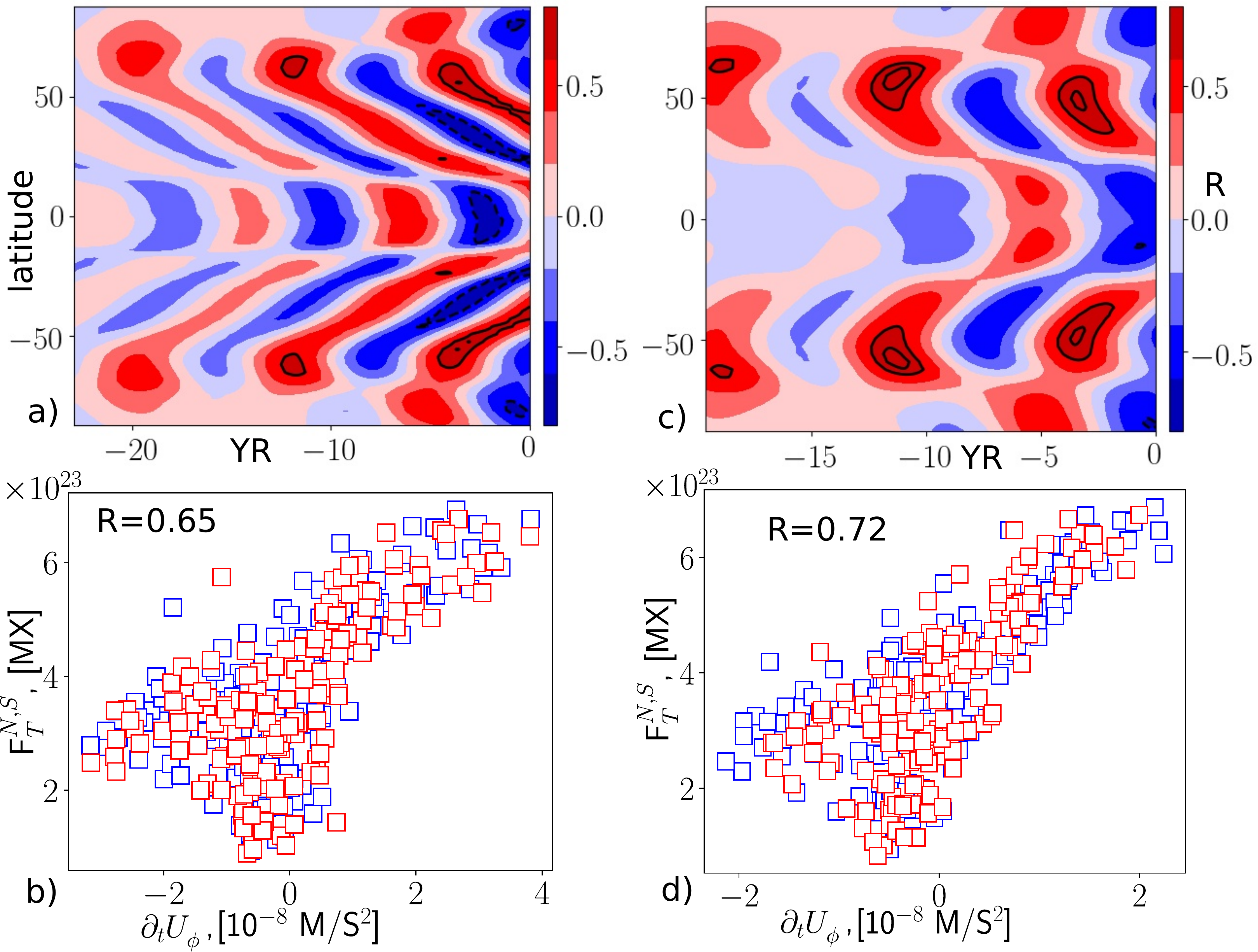}
\caption{\label{fig:CR-C4}{Model C4: a) cross-correlations of the time
series of the total magnetic flux in the subsurface layer 
in the Northern and Southern hemispheres
and parameters of the zonal acceleration, contours show the R-levels
of values $\pm$0.6, $\pm$0.7, and $\pm$0.8; b) the total
toroidal flux in the North (blue squares), $F_{T}^{N}$, and South
$F_{T}^{S}$, (red squares) vs the zonal acceleration $\partial_{t}\overline{U}_{\phi}$ at
$\pm60^{\circ}$ latitudes with the time lag of about 12 years; c) and
d) the same as in panels a) and b) for $r=0.76R_\odot$.}}
\end{figure}

{ To explore correlations of the torsional oscillations with
the magnetic activity, we apply the cross-correlation analysis of time-latitude
variations of the $\partial_{t}\overline{U}_{\phi}$ and magnitudes
of the toroidal magnetic field flux in the North and South hemisphere,
$F_{T}^{N}$ , $F_{T}^{S}$. 

Figures \ref{fig:CR-C4}a and \ref{fig:CR-C4}c show
the time-latitude cross-correlations of the time series of $F_{T}^{(N)}$ and $F_{T}^{(S)}$ with the time series of the zonal acceleration, $\partial_{t}\overline{U}_{\phi}$,
at the top and near the bottom of the convection zone for model C4. We
see that the cross-correlation diagrams as a function of latitude and the lag time reproduce the mean torsional oscillation pattern, including the extended mode. At the surface, the evolution of $\partial_{t}\overline{U}_{\phi}$
at $\pm$60$^{\circ}$ latitudes is $\sim 12$ years ahead of 
$F_{T}^{(N)}$ , $F_{T}^{(S)}$ with the correlation coefficient R$\approx$0.65.
A tighter connection is found between the surface fluxes and  $\partial_{t}\overline{U}_{\phi}$
at the bottom of the convection zone, $r=0.76R_{\odot}$. 
These variations show a much longer impact on the evolution of the $F_{T}$ parameters. We find that the correlation coefficient R$\approx 0.72$ with
a time lag of about 12 years. 

Figures~\ref{fig:CR-C4}b and \ref{fig:CR-C4}d show the scatter plots for the zonal acceleration at $\pm$60$^{\circ}$ latitudes vs the subsurface toroidal fluxes with the 12-year time lag for the individual values during the dynamo cycles and the mean correlation coefficients. For $\partial_{t}\overline{U}_{\phi}$
at r=0.76$R_{\odot}$, in addition to the correlation with the
flux parameters for the 12-year lag, there is a considerable correlation  for the time lag of about 19 years. The latter is
equal to the mean period of the extended cycle in model C4. The
magnitude of this correlation is R$\approx$0.63. The correlations for the spectral coefficients $U_{,t}^{(\ell)}$ and $F_{T}$ have similar values.}

\begin{figure}\centering
\includegraphics[width=0.7\textwidth]{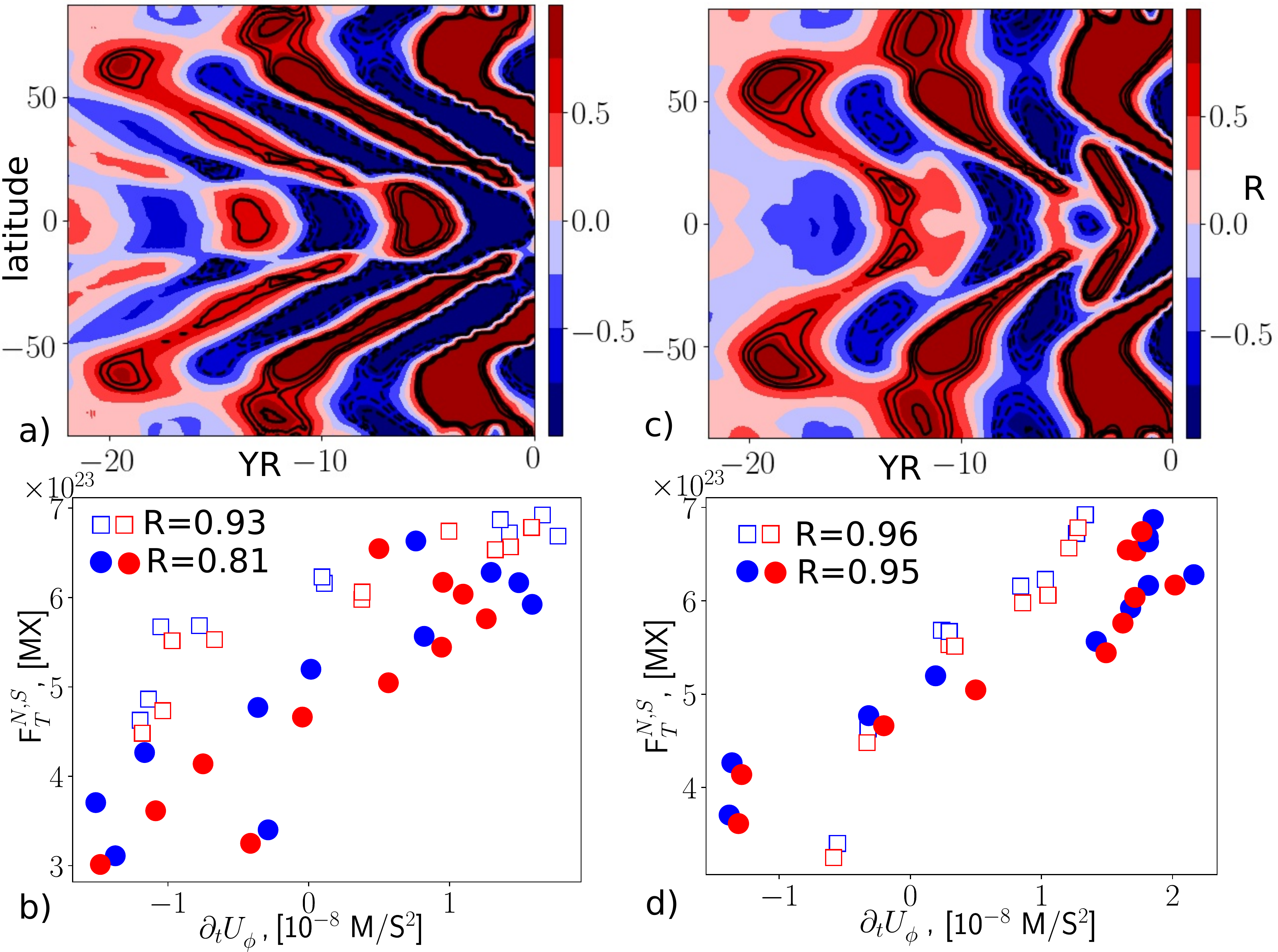}
\caption{\label{fig:CR-C4-A} {Model C4 (with the `centennial' activity variations): a) cross-correlations of the time
series of the total magnetic flux in the subsurface layer in the Northern and
Southern hemispheres and parameters of the zonal acceleration, contours
show the R-levels of values $\pm$0.6, $\pm$0.7, and $\pm$0.8; b)
the total toroidal flux in the North (blue symbols), $F_{T}^{N}$,
and South $F_{T}^{S}$, (red symbols) vs zonal acceleration $\partial_{t}\overline{U}_{\phi}$
at $\pm60^{\circ}$ latitudes with the time lag of about 12 years;
c) and d) the same as in panels a) and b) for the zonal acceleration at $r=0.76R_{\odot}$. Squares show the cycle properties during the periods of the `centennial' rise of magnetic activity, full circles - during the decline periods.}}
\end{figure}

\begin{figure}\centering
	\includegraphics[width=0.7\textwidth]{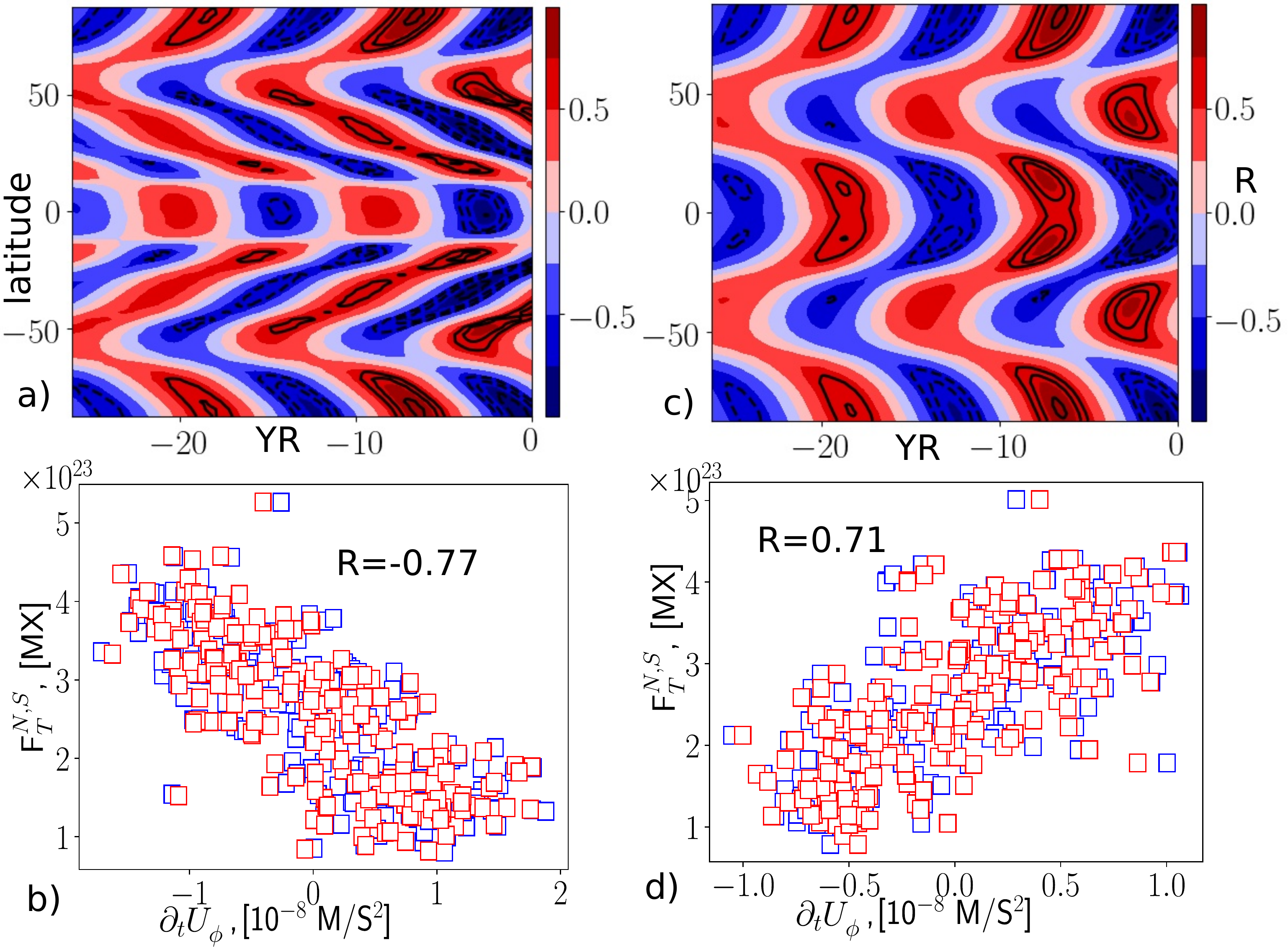}
	\caption{\label{fig:CR}  {The same as in Figure \ref{fig:CR-C4} for
model C5. Note that panel d) shows the correlation of the polar variations
of $\partial_{t}\overline{U}_{\phi}$ at r=0.76R$_{\odot}$  and the maxima of $F_{T}^{N}$,
$F_{T}^{S}$ with a time lag of 19 years. }}
\end{figure}

\begin{figure}\centering
	\includegraphics[width=0.7\textwidth]{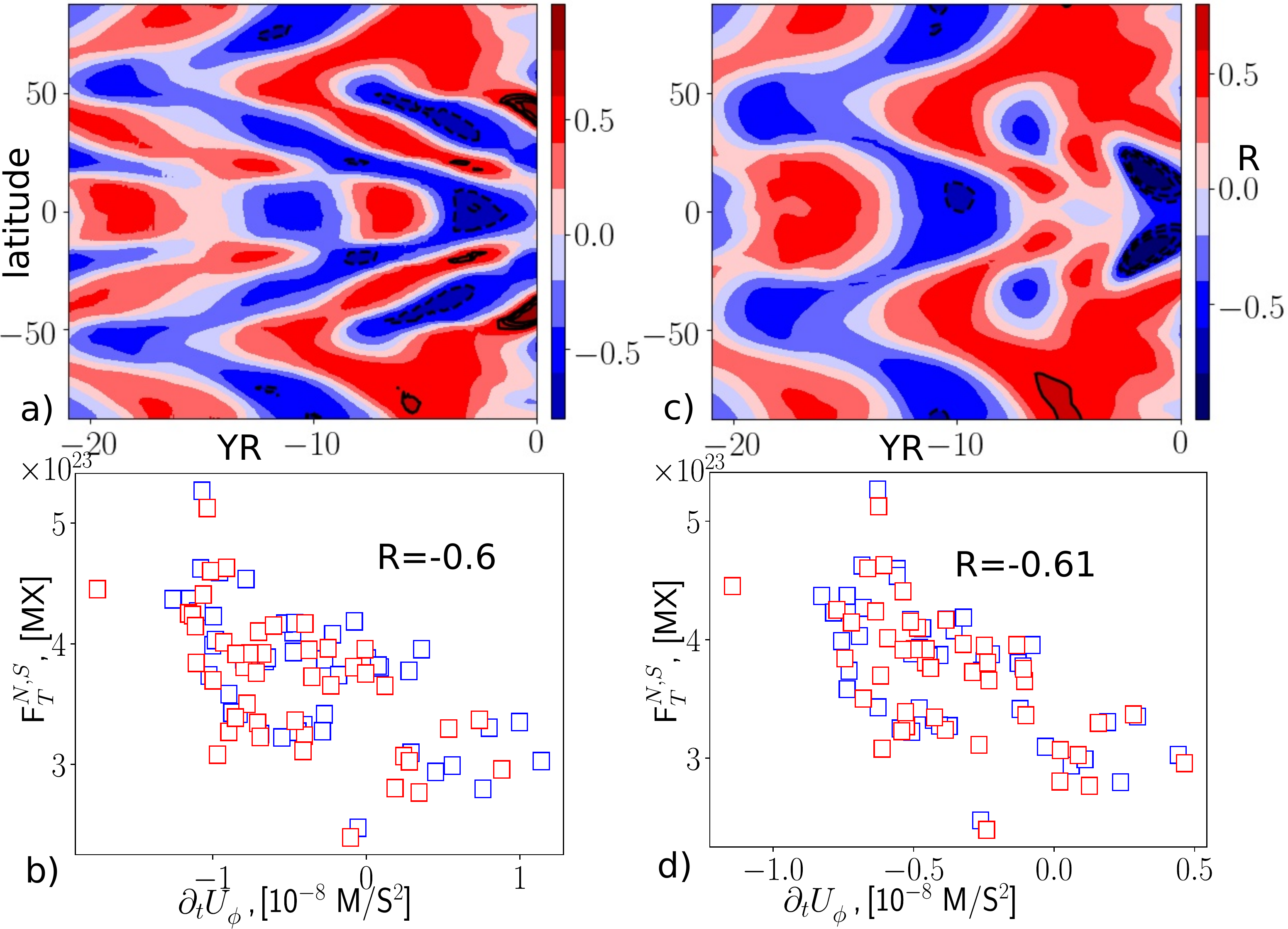}
	\caption{\label{CR-A}The same as in Figure~\ref{fig:CR-C4-A} for model C5 with random variations of the activity cycles.}
\end{figure}

{Also, we study the cross-correlation between continuous time-series of $\partial_{t}\overline{U}_{\phi}$ and the maxima of $F_{T}^{N}$, $F_{T}^{S}$ as a function of the time lag and latitude. 
The results for model C4 are shown in Figure \ref{fig:CR-C4-A}.
We find that, in the subsurface layer, the polar variations of the $\partial_{t}\overline{U}_{\phi}$
correlate with following maxima of the surface fluxes $F_{T}^{(N)}$ , $F_{T}^{(S)}$ with the time lag of about 12 years, with the correlation coefficient R=0.82. Besides, the magnitude of the extended
mode of $\partial_{t}\overline{U}_{\phi}$ at the latitudes $\pm$60$^{\circ}$
correlates with the $F_{T}^{(N)}$ , $F_{T}^{(S)}$ maxima
with the time lag of 19 years, R=0.78. The variations of $\partial_{t}\overline{U}_{\phi}$ at the base of the convection zone at $r=0.76R_\odot$ show a higher level correlations of R=0.88 for the same time lag. Also, results
in Figure \ref{fig:CR-C4-A}b) and d) show that the correlation coefficients
are higher during the centennial rise of the magnetic activity (squares) than those during the centennial decline (full circles). Similar conclusions can be drawn about correlations
for the spectral coefficients $U_{,t}^{(\ell)}$ both for the top
and the bottom of the convection zone.}

{We note that, in model C4, the correlations of the flow parameters
with the maxima of the total magnetic flux of the following cycles are higher than the correlations with the continuous time-series of the flux $F_{T}$. The opposite situation is found for model C5. The results of this model are shown in
Figures \ref{fig:CR} and \ref{CR-A}. We find that the continuous
time-series of $F_{T}$ have the correlation coefficient R$\approx$0.77 for the 11-year time lag the polar variations of $\partial_{t}\overline{U}_{\phi}$ both  at the surface and at $r=0.76R_\odot$. 
We find that the correlation coefficients of the $F_{T}$ maxima correlate with the polar variations $\partial_{t}\overline{U}_{\phi}$ for the 11-year time lag 
are R$\approx$-0.6 at the surface and at $r=0.76R_{\odot}$.}

\section{Discussion and Conclusions}

In the study, using the solar-type non-linear dynamo models, we explored
relationships between the torsional oscillations and the magnetic
activity cycles and investigated the potential of these relationships
for magnetic cycle forecasting. The idea is not new; for example,
\citet{Yoshimura1993A} found evidence that variations in the total
angular momentum on the surface were ahead of the centennial variations
in the magnetic activity of the Sun. A similar effect was found in
models of \citet{Knobloch1998} and \citet{Pipin1999}. Based on the helioseismic analysis of the evolution of the torsional oscillations in the convection zone, \citet{Kosovichev2019} argued that the magnitude of the zonal deceleration in the high-latitude region (at $\sim 60^\circ$ latitude) at the base of the convection zone (at $\sim 0.76R_\odot$) during a solar maximum may provide information about the following solar maximum, that is about 11 years ahead. However, the helioseismology observations cover only the last two solar cycles, so this relationship cannot be verified observationally.

In this paper, we used the previously investigated the relationships between properties of the torsional oscillations at the top and bottom of the convection zone and the surface toroidal magnetic flux characterizing the strength of the activity cycles.
The torsional oscillations in our model are
caused by the influence of the dynamo generated large-scale magnetic
field on the heat transport and turbulent angular momentum fluxes
inside the convection zone, and reproduce the extended mode 
(see, \citealp{Pipin2019c}).
The extended mode of the torsional oscillations covers a period of
the full 22-year magnetic cycle. It was found that at the surface
the maximum of the spectrum of the torsional variations, represented
by the azimuthal velocity acceleration, $\partial_{t}\overline{U}_{\phi}$,
corresponds to the relatively high order harmonic $\ell=9$. Consequently,
the spectrum of $\partial_{t}\Omega=\partial_{t}\overline{U}_{\phi}/r\sin\theta$
has the maximum for $\ell=8$. Our models do not show the extended
wave of torsional oscillations at the bottom of the convection zone
where the spectral maximum is at $\ell=3$. 

{In our study, we identified several precursors which can be
further elaborated for the problems of the solar cycle forecasts.
The cross-correlation analysis shows that the choice of the zonal 
acceleration precursors of the magnetic activity cycles and the type of correlation
analysis can depend on the nature of the magnetic cycle variations.
For the magnetic cycles during the long-term centennial
variation, the best precursors can be found in the analysis of the
temporal relations of the $\partial_{t}\overline{U}_{\phi}$ variations
with the cycle maxima of the magnetic flux parameters. This is demonstrated
by the results of model C4. We note that solar magnetic activity also
tends to show the centennial magnetic cycles \citet{Usoskin2013}, sometimes
called the Gleissberg cycles \citep{Feynman2014}. Model
C4 shows that for this type of the magnetic variations we can expect
a high correlation of the polar variations of the zonal acceleration
with the magnetic flux parameters both for the surface flows and for
the flows near the bottom of the convection zone at r=0.76$R_{\ensuremath{\odot}}$,
where variations of the zonal acceleration show correlations
with the maxima of the subsurface toroidal magnetic flux with the correlation coefficient R=0.93 with the time lag of about 12 and 19 years. }

{For model C5, which simulates random fluctuations of the cycle magnitude, the best
correlations are found between the continuous time-series of the zonal acceleration and the toroidal flux. In the best case, at the surface, the continuous time-series of the toroidal flux correlates with the polar variations of the zonal acceleration
with R$\approx$-0.77 for the time lag around 12 years. In agreement
with the model C4 results, the higher R is found for shorter 
time-series on the long-term growth or descend  
of the magnetic activity. Both dynamo models reproduce the correlation between the dipole moment and the toroidal flux maximum for the time lag of 5-6 years (with R=0.7-0.9), which is currently considered as the most robust relationship for the cycle prediction.}

Therefore, we conclude that these parameters of torsional oscillation
have a  longer forecast horizon than the predictions made with the help
of the dipole components of the poloidal magnetic field of the Sun. { The model results show that for improving prediction
of the solar cycles, it is essential to consider the continuous evolution
of the cycle properties, and not only their values during the cycle
extrema. In fact, the continuous evolution of solar properties for 
the cycle prediction had been studied in the past. For example,
\citet{Makarov1989} studied a correlation of the polar faculae activity
and the evolution of the Wolf's sunspot number parameter.  The data assimilation
in the dynamo models \citep[e.g.][]{Kitiashvili2010,Dikpati2016b,Hung2017} provides 
a systematic approach to employ the continuous evolution of  
observed properties for solar cycle forecasting.} 

We checked these relationships for other dynamo models without the
extended mode, such as model M7 from PK19, in which the influence
of the magnetic field on the heat transport was neglected. We did
not find the same precursors of the torsional oscillations in that
model. We conclude that the extended mode of the torsional oscillations
is crucial for the cycle predictions based on flow characteristics. 
This conclusion is in general agreement with the suggestion made by
\citet{Kosovichev2019} based on a helioseismic analysis of the torsional
oscillations.

{In summary, it is found that the torsional oscillations parameters, including  the extended 22-yr mode show a considerable correlation with subsequent cycle magnitudes for the time lag in the range of  11-20 yr. The sign of correlation and the time-delay parameters can depend on the properties of the long-term variations of the dynamo cycle. 
This theoretical study should be extended using the available observations.}

{Acknowledgments} VP thanks the support of RFBR under grant
19-02-53045, the project II.16.3 of ISTP SB RAS; AK thanks the support
of NASA grants: NNX14AB70G and 80NSSC20K0602.

\bibliographystyle{aasjournal}

\begin{thebibliography}{}
\expandafter\ifx\csname natexlab\endcsname\relax\def\natexlab#1{#1}\fi
\providecommand{\url}[1]{\href{#1}{#1}}
\providecommand{\dodoi}[1]{doi:~\href{http://doi.org/#1}{\nolinkurl{#1}}}
\providecommand{\doeprint}[1]{\href{http://ascl.net/#1}{\nolinkurl{http://ascl.net/#1}}}
\providecommand{\doarXiv}[1]{\href{https://arxiv.org/abs/#1}{\nolinkurl{https://arxiv.org/abs/#1}}}

\bibitem[{{Altrock}(1997)}]{Altrock1997}
{Altrock}, R.~C. 1997, \solphys, 170, 411, \dodoi{10.1023/A:1004958900477}

\bibitem[{{Babcock}(1961)}]{Babcock1961}
{Babcock}, H.~W. 1961, \apj, 133, 572, \dodoi{10.1086/147060}

\bibitem[{{Brandenburg}(2018)}]{Brandenburg2018}
{Brandenburg}, A. 2018, Journal of Plasma Physics, 84, 735840404,
  \dodoi{10.1017/S0022377818000806}

\bibitem[{{Brandenburg} \& {Subramanian}(2005)}]{Brandenburg2005b}
{Brandenburg}, A., \& {Subramanian}, K. 2005, \physrep, 417, 1,
  \dodoi{10.1016/j.physrep.2005.06.005}

\bibitem[{{Cameron} \& {Sch{\"u}ssler}(2017)}]{Cameron17}
{Cameron}, R.~H., \& {Sch{\"u}ssler}, M. 2017, \aap, 599, A52,
  \dodoi{10.1051/0004-6361/201629746}

\bibitem[{{Charbonneau}(2011)}]{Charbonneau2011}
{Charbonneau}, P. 2011, Living Reviews in Solar Physics, 2, 2

\bibitem[{{Choudhuri} {et~al.}(2007){Choudhuri}, {Chatterjee}, \&
  {Jiang}}]{Choudhuri2007}
{Choudhuri}, A.~R., {Chatterjee}, P., \& {Jiang}, J. 2007, Physical Review
  Letters, 98, 131103, \dodoi{10.1103/PhysRevLett.98.131103}

\bibitem[{{Choudhuri} \& {Dikpati}(1999)}]{Choudhuri1999}
{Choudhuri}, A.~R., \& {Dikpati}, M. 1999, \solphys, 184, 61

\bibitem[{{Dikpati} {et~al.}(2016){Dikpati}, {Anderson}, \&
  {Mitra}}]{Dikpati2016b}
{Dikpati}, M., {Anderson}, J.~L., \& {Mitra}, D. 2016, \apj, 828, 91,
  \dodoi{10.3847/0004-637X/828/2/91}

\bibitem[{{Feynman} \& {Ruzmaikin}(2014)}]{Feynman2014}
{Feynman}, J., \& {Ruzmaikin}, A. 2014, Journal of Geophysical Research (Space
  Physics), 119, 6027, \dodoi{10.1002/2013JA019478}

\bibitem[{{Hung} {et~al.}(2017){Hung}, {Brun}, {Fournier}, {Jouve},
  {Talagrand}, \& {Zakari}}]{Hung2017}
{Hung}, C.~P., {Brun}, A.~S., {Fournier}, A., {et~al.} 2017, \apj, 849, 160,
  \dodoi{10.3847/1538-4357/aa91d1}

\bibitem[{{Kitchatinov} {et~al.}(2018){Kitchatinov}, {Mordvinov}, \&
  {Nepomnyashchikh}}]{Mordvinov2018}
{Kitchatinov}, L.~L., {Mordvinov}, A.~V., \& {Nepomnyashchikh}, A.~A. 2018,
  \aap, 615, A38, \dodoi{10.1051/0004-6361/201732549}

\bibitem[{{Kitchatinov} \& {R{\"u}diger}(1999)}]{Kitchatinov1999a}
{Kitchatinov}, L.~L., \& {R{\"u}diger}, G. 1999, \aap, 344, 911

\bibitem[{{Kitchatinov} \& {R{\"u}diger}(2005)}]{Kitchatinov2005}
---. 2005, Astronomische Nachrichten, 326, 379, \dodoi{10.1002/asna.200510368}

\bibitem[{{Kitiashvili} \& {Kosovichev}(2010)}]{Kitiashvili2010}
{Kitiashvili}, I.~N., \& {Kosovichev}, A.~G. 2010, in IAU Symposium, Vol. 264,
  IAU Symposium, ed. .~J.-P.~R. A.~G.~Kosovichev, A.~H.~Andrei, 202--209,
  \dodoi{10.1017/S1743921309992638}

\bibitem[{{Kleeorin} {et~al.}(2000){Kleeorin}, {Moss}, {Rogachevskii}, \&
  {Sokoloff}}]{Kleeorin2000}
{Kleeorin}, N., {Moss}, D., {Rogachevskii}, I., \& {Sokoloff}, D. 2000, \aap,
  361, L5

\bibitem[{Kleeorin \& Rogachevskii(1999)}]{Kleeorin1999}
Kleeorin, N., \& Rogachevskii, I. 1999, Phys. Rev.E, 59, 6724

\bibitem[{{Knobloch} {et~al.}(1998){Knobloch}, {Tobias}, \&
  {Weiss}}]{Knobloch1998}
{Knobloch}, E., {Tobias}, S.~M., \& {Weiss}, N.~O. 1998, \mnras, 297, 1123,
  \dodoi{10.1046/j.1365-8711.1998.01572.x}

\bibitem[{{Kosovichev} \& {Pipin}(2019)}]{Kosovichev2019}
{Kosovichev}, A.~G., \& {Pipin}, V.~V. 2019, \apj, 871, L20,
  \dodoi{10.3847/2041-8213/aafe82}

\bibitem[{Krause \& R\"adler(1980)}]{Krause1980}
Krause, F., \& R\"adler, K.-H. 1980, Mean-Field Magnetohydrodynamics and Dynamo
  Theory (Berlin: Akademie-Verlag), 271

\bibitem[{{Makarov} {et~al.}(1989){Makarov}, {Makarova}, \&
  {Sivaraman}}]{Makarov1989}
{Makarov}, V.~I., {Makarova}, V.~V., \& {Sivaraman}, K.~R. 1989, \solphys, 119,
  45, \dodoi{10.1007/BF00146211}

\bibitem[{{Mitra} {et~al.}(2010){Mitra}, {Candelaresi}, {Chatterjee},
  {Tavakol}, \& {Brandenburg}}]{Mitra2010}
{Mitra}, D., {Candelaresi}, S., {Chatterjee}, P., {Tavakol}, R., \&
  {Brandenburg}, A. 2010, Astronomische Nachrichten, 331, 130,
  \dodoi{10.1002/asna.200911308}

\bibitem[{Parker(1955)}]{Parker1955}
Parker, E. 1955, Astrophys. J., 122, 293

\bibitem[{{Parker}(1984)}]{Parker1984}
{Parker}, E.~N. 1984, \apj, 281, 839, \dodoi{10.1086/162163}

\bibitem[{{Paxton} {et~al.}(2011){Paxton}, {Bildsten}, {Dotter}, {Herwig},
  {Lesaffre}, \& {Timmes}}]{Paxton2011}
{Paxton}, B., {Bildsten}, L., {Dotter}, A., {et~al.} 2011, \apjs, 192, 3,
  \dodoi{10.1088/0067-0049/192/1/3}

\bibitem[{{Paxton} {et~al.}(2013){Paxton}, {Cantiello}, {Arras}, {Bildsten},
  {Brown}, {Dotter}, {Mankovich}, {Montgomery}, {Stello}, {Timmes}, \&
  {Townsend}}]{Paxton2013}
{Paxton}, B., {Cantiello}, M., {Arras}, P., {et~al.} 2013, \apjs, 208, 4,
  \dodoi{10.1088/0067-0049/208/1/4}

\bibitem[{{Pipin}(1999)}]{Pipin1999}
{Pipin}, V.~V. 1999, \aap, 346, 295

\bibitem[{{Pipin}(2018)}]{Pipin2018b}
---. 2018, Journal of Atmospheric and Solar-Terrestrial Physics, 179, 185,
  \dodoi{10.1016/j.jastp.2018.07.010}

\bibitem[{{Pipin} \& {Kitchatinov}(2000)}]{Pipin2000}
{Pipin}, V.~V., \& {Kitchatinov}, L.~L. 2000, Astronomy Reports, 44, 771,
  \dodoi{10.1134/1.1320504}

\bibitem[{{Pipin} \& {Kosovichev}(2018)}]{Pipin2018c}
{Pipin}, V.~V., \& {Kosovichev}, A.~G. 2018, \apj, 854, 67,
  \dodoi{10.3847/1538-4357/aaa759}

\bibitem[{{Pipin} \& {Kosovichev}(2019)}]{Pipin2019c}
---. 2019, \apj, 887, 215, \dodoi{10.3847/1538-4357/ab5952}

\bibitem[{{Rempel}(2005)}]{Rempel2005c}
{Rempel}, M. 2005, \apj, 631, 1286, \dodoi{10.1086/432610}

\bibitem[{{Schatten} {et~al.}(1978){Schatten}, {Scherrer}, {Svalgaard}, \&
  {Wilcox}}]{Schatten1978}
{Schatten}, K.~H., {Scherrer}, P.~H., {Svalgaard}, L., \& {Wilcox}, J.~M. 1978,
  \grl, 5, 411, \dodoi{10.1029/GL005i005p00411}

\bibitem[{{Spruit}(2000)}]{2000SSRv...94..113S}
{Spruit}, H. 2000, \ssr, 94, 113

\bibitem[{{Stix}(1981)}]{Stix1981}
{Stix}, M. 1981, \aap, 93, 339

\bibitem[{{Ulrich} \& {Boyden}(2005)}]{Ulrich2005}
{Ulrich}, R.~K., \& {Boyden}, J.~E. 2005, \apjl, 620, L123,
  \dodoi{10.1086/428724}

\bibitem[{Usoskin(2013)}]{Usoskin2013}
Usoskin, I.~G. 2013, Living Reviews in Solar Physics, 10, 1,
  \dodoi{10.12942/lrsp-2013-1}

\bibitem[{{Yoshimura} \& {Kambry}(1993)}]{Yoshimura1993A}
{Yoshimura}, H., \& {Kambry}, M.~A. 1993, Astronomische Nachrichten, 314, 9,
  \dodoi{10.1002/asna.2113140104}

\end{thebibliography}

\end{document}